# A Case Study to Identify the Hindrances to Widespread Adoption of Electric Vehicles in Qatar

Amith Khandakar [1], Annaufal Rizqullah [1], Anas Ashraf Abdou Berbar [1], Mohammad Rafi Ahmed [1], Atif Iqbal [1], Muhammad E. H. Chowdhury [1] and S. M. Ashfaq Uz Zaman [2,*]

[1] Department of Electrical Engineering, Qatar University, PO BOX 2713, Doha, Qatar; amitk@qu.edu.qa (A.K.); ar1605250@student.qu.edu.qa (A.R.); ab1605615@student.qu.edu.qa (A.A.A.B.); ma1603851@student.qu.edu.qa (M.R.A.); atif.iqbal@qu.edu.qa (A.I.); mchowdhury@qu.edu.qa (M.E.H.C.)
[2] Qatar Emiri Naval Forces, Gulf Arabian, PO BOX 2237, Doha, Qatar
\* Correspondence: ashfaquzzaman2@gmail.com; Tel.: +974-5568-2068

**Abstract:** The adoption of electric vehicles (EVs) have proven to be a crucial factor to decreasing the emission of greenhouse gases (GHG) into the atmosphere. However, there are various hurdles that impede people from purchasing EVs. For example, long charging time, short driving range, cost and insufficient charging infrastructures available, etc. This article reports the public perception of EV-adoption using statistical analyses and proposes some recommendations for improving EV-adoption in Qatar. User perspectives on EV-adoption barriers in Qatar were investigated based on survey questionnaires. The survey questionnaires were based on similar studies done in other regions of the world. The study attempted to look at different perspectives of the adoption of EV, when asked to a person who is aware of EVs (technical respondents—people studying/working at universities/research centers and policy makers) or a person who may or may not be aware of EVs (non-technical respondents—people working in banks, governments and private non-academic organizations, etc.). Cumulative survey responses from the two groups were compared and analyzed using two-sample *t*-test statistical analysis. Detailed analyses showed that—among various major hindrances—raising of public awareness of such greener modes of transportation, the availability of charging options in more places and policy incentives towards EVs would play a major role in EV-adoption. The authors provide recommendations that—along with government incentives—could help make a gradual shift to a greater number of EVs convenient for people of Qatar. The proposed systematic approach for such a study and analysis may help in streamlining research on policies, infrastructures and technologies for efficient penetration of EVs in Qatar.

## 1. Introduction

The world population is growing constantly. This means that energy consumption from the burning of fossil fuels is increasing every year. Therefore, the emission of the greenhouse gases (GHG) is also increasing at a high rate [1]. Greenhouse gases (GHG), such as carbon dioxide, methane and nitrous oxide trap some of the energy that goes out of the Earth and heat the atmosphere. This causes the Earth's energy received by the sun—and the energy going out from the Earth—to be unbalanced, leading to climate change. climate change is a major problem because it causes catastrophes such as heat waves and floods as well as increases the risk of heat-related illnesses to people [2]. The emission of GHGs into the atmosphere also increases the concentration of air pollution; it has been proven that air pollution is one of the main causes of respiratory diseases. High concentration of pollution in the air has increased the number of people with cardiovascular illnesses, asthma and cancer [3].

Due to its big dependence on fossil fuels consumption, the transportation sector plays a great role in the distribution of greenhouse gases. Research and studies have been done to minimize the burning of fossil fuels and move towards the adoption of renewable [4–7] energies/green energy strategies. The adoption of renewable energies in the transportation sector can be enhanced by reducing the number of internal combustion engine (ICE) vehicles—and by the adoption of EVs. The mass production of the first generation of EVs happened in the 19th century, but these vehicles were ruled out of the market due to their high initial purchase costs and low autonomy. Today, people still prefer purchasing ICE vehicles compared to EVs due to their higher reliability. Over time, the usage of large number of ICE vehicles has led the transportation sector to being one of the top emitters of GHGs. This concerning fact has renewed interest in replacing ICE with EVs [8]. The Republic of China has the largest market of EVs, followed by Europe and the US. The global sales of EVs increased by 2 million from 2017 to 2018, where the total number of EVs purchased in 2018 were just over 5 million. The increasing trend in the purchase of EVs has made the leading countries in EV sales to develop economic instruments that would reduce the differences in cost between EVs and ICE vehicles and develop more charging infrastructures. As technology advances, EV batteries are being further improved, and the manufactures of EVs are being expanded to attract more EV purchases. Furthermore, innovative designs of EVs and better batteries are developed to accelerate the adoption of EV. In the new policies scenario in 2030, it is mentioned that the sales in EVs are predicted to reach 23 million, with over 130 million in stock. In the EV30@30 scenario, EV sales are expected to reach 43 million in 2030 with a stock of more than 250 million. The new policies scenario is expected to cut the demand for oil by 127 million tons of oil equivalent (Mtoe) which is equal to about 2.5 million barrels per day (mb/d) in 2030. On the other hand, the EVs in the EV30@30 Scenario are expected to reduce oil demand by 4.3 mb/d. The demand of electricity to serve the EVs in 2030 in the new policies scenario is calculated to be almost 640 terawatt hours (TWh) while that of the EV30@30 Scenario is 110 TWh. By using well-to-wheel technologies, the GHGs emitted from EVs will continue to be reduced compared to emissions from ICE vehicles. GHG emission from the EVs as per the new policies scenario will reach almost 230 million tons of carbon-dioxide equivalents (Mt $CO_2$-eq) in 2030. This value will compensate about 220 Mt $CO_2$-eq emissions. An average using electricity that is characterized by the current global average carbon intensity which is 518 g of carbon dioxide equivalent per kilowatt-hour (g $CO_2$-eq/kWh) emits lower GHG than an average ICE vehicle that uses gasoline for its whole lifetime. However, the reduced emission of $CO_2$ is more effective for hybrid vehicles, compared to EVs in countries where power generation is obtained from the burning of coal [9].

The adoption rate of EVs still varies around the world. However, this adoption can be affected by the vehicle price, total cost of owning EVs, driving experience, the availability of Charging Station (CS), social influence, environmental awareness and others [10]. In [11], it is stated that the popular EV categories: i) battery-based EVs (BEVs)—complete EVs without any ICE options, and ii) plug in hybrid EV (PHEV)—EVs that have an ICE option and high storage capacity battery with options of charging, has increased significantly from 2010 to 2016, as shown in Figure 1. Figure 1 also shows the anticipated trends of adoption of EVs until 2030 using the data available from [9].

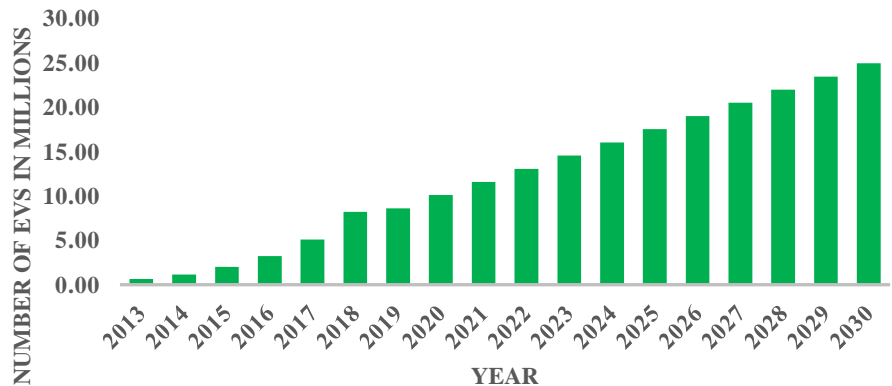

**Figure 1.** Trend of electric vehicles (EVs) in the global market [9]. Reproduced from [9], Global EV outlook: 2019.

Although EVs provide many benefits to the environment, there are still a lot of barriers that prevent them from be widely adopted around the world. Many studies report barriers to EV-adoption in different parts of the world [12–15]. According to [12], consumer viewpoint is one of the most significant barriers that limits the widespread adoption of EV. The authors focused on three main aspects that fall under the consumers demand: financial, performance and the infrastructure. Financial barriers consist of the price of EVs, cost of the battery, lack of knowledge of the fuel cost and maintenance cost. The initial purchase cost comparison between ICE vehicles and EVs from different manufactures are tabulated in Table 1 [16–26]. For each manufacturer, an EV is compared with an equivalent ICE vehicle of the same manufacturer in terms of model, category and size. It can be seen that the buying price of EVs and PHEVs is exorbitant compared to ICE vehicles. In contrast, the maintenance and operational costs of EVs are much lower than ICE vehicles. This comparison is shown in Table 2. The maintenance cost comparison mainly considers the first 100,000 miles of a vehicle [27,28]. Table 2 shows the basic costs of maintaining an EV or ICE vehicle. Many other options and safety options are offered by vendors that are not considered in this table. Operating cost is the main continuous cost involved in operating vehicles. While EVs consume electricity, ICE vehicles consume gasoline. The prices of electricity and gasoline are different in each country. In Qatar, the rates of electricity are USD 0.032/kWh and the average gasoline price in 2020 is USD 1.65/gallon. A large fluctuation in gasoline price has been seen due to COVID-19; however, this can be considered an outlier, due to this extraordinary situation not seen in the last century. The EV Nissan Leaf has a 40 kWh battery and a range of 149 miles [16–26]. According to the electricity rates in Qatar, it would cost 0.0086$/mile. The ICE vehicle Honda Civic has tank capacity of 12.4 gallons and a consumption of 30 miles/gallon. With Qatar gasoline average rates, it would cost USD 0.055/mile. It is clear that an EV has around 5–6 times cheaper operating costs than ICE vehicles: For 100,000 miles, an average operating cost for Nissan Leaf is USD 860 and for Honda Civic is USD 5500 [16–26]. EV performance barriers also consist of the safety, reliability, range, the life of the EV battery, the time for an EV to charge and the power of the vehicle. Finally, infrastructure barriers consist of the availability of the CS in various places such as highways, workplaces and public places. In [12], it was also mentioned that consumer characteristics such as the gender, age, income and environmental awareness play a major role in the adoption of EVs. For example, people who

are highly educated and aware of the environmental benefits of EVs have a higher likelihood of choosing an EV over an ICE vehicle.

**Table 1.** Cost comparison between EVs and Internal Combustion Engine Vehicles (ICEV) [Price as of May 2020].

| Manufacturer | Model | Year | Type | Price |
|---|---|---|---|---|
| KIA | Niro EV | 2020 | EV | USD 38,500 |
| | Niro | 2020 | ICEV | USD 24,590 |
| Volkswagen | ID.3 | 2020 | EV | USD 37,321 |
| | Golf | 2020 | ICEV | USD 23,195 |
| Nissan | Leaf Hatchback | 2020 | EV | USD 31,600 |
| | Centra | 2020 | ICEV | USD 19,090 |
| Chevrolet | Bolt | 2020 | EV | USD 37,495 |
| | Spark | 2020 | ICEV | USD 18,595 |
| Hyundai | IONIQ | 2020 | PHEV | USD 35,763 |
| | i30 | 2020 | ICEV | USD 26,605 |

**Table 2.** Maintenance cost comparison.

| Property | EVs Cost | ICE vehicles Cost |
|---|---|---|
| Oil change | 0 | $600 |
| Automatic transmission fluid | 0 | $70 |
| Sparks plugs and wires | 0 | $200 |
| Muffler replacement | 0 | $150 |
| Brakes maintenance | USD 200 | $400 |
| Timing belt and water pump | 0 | $1000 |
| Total | USD 200 | $2420 |

Developing sustainable strategies for using energy resources efficiently requires a thorough evaluation of the three main factors of sustainability, i.e., society, economy and the environment. The widespread EV adoption is expected to change the structure of the energy industry. This change in the structure can be from the employment shifts in the supply chain of different energy sectors to tax balances and profitability. The utilization of renewable energies can be an important strategy to maximize the environmental benefits from the adoption of EVs. These environmental benefits are highly dependent on the source of electricity generation. For example, use of EVs in countries that depend on petroleum for generating electric power could be lower than using ICE vehicles with the exception of Norway which have very high penetration of EVs, due to the successful strategies it follows and should be used as an example by others. EV technologies are environmentally beneficial since they have the potential to minimize air pollution, airborne diseases, climate change, energy consumption and water footprint. According to International Energy Agency (IEA), some countries have set a target on the number of EVs they will adopt; some other countries have banned ICE vehicles. In the Middle East region, Qatar have targeted a 10% EV sales by the year 2030. The Qatar national vision 2030 is aiming to increase the economic and social growth as well as improving the environment. Qatar is one of the largest $CO_2$ emitters per capita in the world [29], which has encouraged Qatar to reduce $CO_2$ emissions. While the major power generation in Qatar relies on natural gas, the country is currently constructing a large-scale solar power plant that will have a capacity of 350 MW [30]. According to World Health Organization (WHO), Doha is known to be the twelfth-most-polluted city in the world, .and that the high concentration of pollution in the city poses major health risks to people [29]. In this study, the authors studied the transportation system of Qatar and the factors that would help the EV-adoptions in Qatar.

Solar and wind power in Qatar may be the most significant renewable energy sources to compensate the usage of natural gases. An initiative called "green car" that aims to achieve sustainable transportation in Qatar has been signed by the Ministry of Energy and Industry, Ministry of Transport and Communications and KAHRAMAA. This initiative aims to increase the total number of EVs to a minimum of 10%—and to provide the CS required by these vehicles [29].

Several regional research works were carried out by different researcher in [9,12–14,31–33] but the studies do not reflect the possible reasons—according to public perception—as barriers for EV-adoption in a financially stable Middle Eastern country like Qatar. Moreover, the articles do not provide the details of the surveys. Furthermore, detailed analyses of the survey results is not provided in the articles. All these points are addressed in this study. The contribution of this study could be stated as: (i) systematic design of the survey to get two different public perception (technical and non-technical, which will be discussed in details later) of the EV-adoption, (ii) survey result analysis using statistical methods such as two-sample *t*-test analysis [34,35], (iii) recommendation based on the survey analysis.

This study is organized as follows: Section 2 shows the considerations of adopting EVs in the literatures. Section 3 presents an overview of the implemented study to check the barriers of EV-adoption in mass scale at Qatar. The section also provides details of the survey conducted, along with how the questions were selected to get meaningful results. It also provides details of statistical analysis to interpret the survey results. Section 4 provides analysis carried out with survey followed by Section 5 where the survey results and statistical analysis results are provided. This section also provides valuable recommendation to help in EV-adoption based on the survey results. Finally, Section 6 presents conclusions and summarizes suggestions from the authors for improvements.

## 2. Considerations in EV Adoption in Literatures

The authors in [12] conducted a survey from the drivers in Tianjin, China, to study the viewpoint of the consumers on EVs. The questions are developed and divided into four parts: (i) General information such as gender, age, income, level of education, etc.; (ii) Fourteen possible barriers to rate; (iii) Respondents were asked to rate the policy incentives that may enhance the widespread of EVs such as paying fewer taxes or free parking for EVs; and (iv) Three questions were asked to the respondents about their willingness to purchase EVs, recommendation of EVs to others and having more EVs in the market. Each question is rated on a scale of 1 to 5 where 1 is 'strongly unwilling' and 5 is 'strongly willing'. Once the questionnaire was done, the authors ensured that the quality of the survey is high by removing the invalid questionnaires. The questionnaires are said to be invalid if the respondents completed it in less than 6 min, and/or there are missing answers, and/or there are six identical consecutive answers. Online questionnaires were also conducted to collect more samples for the study of the consumer behavior towards EVs. The survey answers were analyzed using the chi-squared test to explore the different perceptions of people on EVs.

A study of EV-adoption was also conducted in Bangladesh [14], which showed that the adoption of EVs would be beneficial to the environment in Bangladesh, but that the increase in electricity demand may produce harmonics in feeder current, hence reducing power quality. Moreover, the power losses in the distribution transformers and voltage disturbances would become major problems. The authors in [14] emphasized on some problems that may occur in the adoption of EVs in Bangladesh. These are lack of EVs CS, battery charging affects power quality, shortage of power supply, battery price and capacity, high charging cost and time, short EV battery lifetime, low EV speed, frequent accident, lack of government incentives, low driving range and uneven road tracks. The Bangladesh government has legalized 100,000 electric vehicles to operate on the rural roads in 2019 and the Bangladesh Road Transport Authority (BRTA) plans to bring these vehicles under their oversight. To make the adoption of these EVs more convenient, the Bangladesh government has installed eight 30-kW solar charging stations in the country which can charge 20 EVs at a time [36]. Developing countries like Bangladesh are not predicted to have a huge sales of EVs until the next decade, since the government had pledged that all new cars that will be sold will be electric by 2030 [37]. In comparison to the data for motor vehicle sales in Bangladesh between 2005 to 2019 which has an average of 38,500 units per year, there is a potential for Bangladesh to bring in over 30 thousand EVs per year by 2030 if all new vehicles will be electric [37]. The potential rise in the number of electric vehicles in Bangladesh will require more charging infrastructures and other facilities for a more convenient adoption of EVs. By the end of 2019, there were a total of 504,130 registered vehicles in Bangladesh—of which around 100,000 of these are electric. If most people start switching from combustion engine vehicles to electric vehicles, Bangladesh may reach over 300,000 EVs across the country [37].

In addition to the above, studies done in other Asian countries that are economically different from Qatar, one case study that was very useful for the research in this study was the example of Norway, also an oil-rich economy. In [31], the authors examined how incentives effect EV-adoption in Norway. Norway is one of the countries that adopted EVs early, studying the incentives of Irish people for EV-adoption. In Norway, EVs have started the growth up since 2009, starting from around 2000 EVs and reaching almost 50,000 EVs in [31]. The market share of EVs in Norway 2020 has broken records in the world. By March 2020, the market share of EVs have reached around 75.2% of vehicles sales. The remaining 25% is divided between plug-less hybrids of 7.1%, diesel of 10% and petrol of 7.7% [38]. Norway has massive amounts of taxes for purchasing new vehicles. One of the main incentives of EV-adoption is the taxes exemption which drops down the prices of EVs to similar or even cheaper than ICE. In addition, EVs are exempted from the value added tax (VAT) which is huge as 25%. One more incentive is the fees of issuing and renewing a vehicle license, EVs have the lowest price in all fees regarding the vehicle license. There are more incentives in the local community of Norway such as free road tolling and parking slots. Lastly, Bus lanes are accessible by BEVs.

A similar study is done in a neighboring country of Qatar in [13], where the availability of infrastructure in Muscat, Oman, were analyzed for convenient accessibility. This analysis was done by looking at how the existing fuel stations are distributed. The distance for driving between different fuel stations, zone activities (e.g., hospitals, markets, housings, schools, etc.) and the knowledge of the people about EVs were also studied. Finding efficient locations for EVs CS may accelerate the adoption of EV, since it reduces the anxiety of the users related to battery recharge [13]. In [13], the city of Muscat was divided into 17 zones and it was found that there was no logical connection between the existing fuel stations and the zones. The conclusion that could be made from the study of the fuel station distribution in [13] is that the area of the zone does not determine the number of stations in this zone, but it is determined by the population and activities existing in the zone. Populated zones and public places such as grocery stores and hospitals tend to have more fuel stations.

Shorter distances between CS will also improve the convenience of EV-charging. It is crucial to know about the maximum driving range of the fully charged EVs, and the CS should be built with respect to the maximum driving range. Since the fuel stations were pretended to be EV-charging stations in [13] and the distance between each station were analyzed, Muscat shows a potential in EV-adoption if the driving range of the vehicles are limited to be within 100–200 km.

There are relevant studies on the adoption of EVs in the Middle Eastern countries. In [39], the authors have discussed the opportunities, challenges faced in the development of smart cities in the countries of Gulf Cooperation Council (GCC). The authors have stressed the need for technological advances of fast EV-charging stations and increasing the driving range to further enhance the adoption of EVs. The authors have provided examples of how well know international companies such as Tesla have tried bringing in some of their technologies in Abu Dhabi and Dubai to promote their use. The study utilizes the study and further strengthens the recommendations using the public perception using survey results.

Similar studies on conducting a survey to analyze the factors hindering the EV-adoption was done in [32,33]. The papers stated that despite government incentives, EV-adoption was not reaching a higher scale. This is due to the buying price, battery performance, driving range and charging time. There are external factors like CS, fuel price, public visibility and social norms. However, EVs have made its way into the market due to the development of battery technology, vehicles efficiency and due to less air pollution. Considering these benefits, several countries started setting targets in EV-adoption. According to the results of the surveys in these papers, vehicles price and its characteristics plays a major role in the decision of purchasing an EV. One of the survey results implied that 55% of the respondent marked "major disadvantage" and 30% marked "disadvantage" on EV-adoption due to its higher purchasing, i.e., people are not willing to pay higher price which an EV demands. On the other hand, industry survey showed that 63% of buyer claims that price of an EV is a large barrier to EV-adoption. As compensation to this, some automotive company reduced the prices of EVs to a significant amount. However, the study suggests that electric power in vehicles provides lower expense in the long run, but not high enough to compensate the higher purchasing price. The literature also found that according the claims of 70% of the respondents in a survey, the driving range of the EV is another major drawback to its adoption. The survey was taken from people living in urban areas, those who are not much concerned about driving ranges. In addition, 33% of the respondents pointed at battery range of an EV as a drawback to its adoption. One of the other surveys suggested that EV consumers are willing to pay extra money to increase the driving range of their EV. The study also showed that 70% of respondent expectation of vehicles befits the specification of a PHEV over BEV. Therefore, the conclusion was made that to eradicate the driving range anxiety, more CS should be built because the charging time of an EV and its driving range are important to consumers of EV. The study also suggested that EV-adoption will be feasible if it can withstand longer driving range with very less charging time. Otherwise, the EV being an efficient vehicle does not add much weight to its adoption. Other studies showed that highly educated people are more likely to prefer an EV rather than a rich person who can easily afford an EV. Sometimes, a person owning multiple cars would not be interested to purchase an EV. However, home-charging infrastructure draws people's interest to EV-adoption. Therefore, people having better educational background being more environments cautious as well as previous EV owners would be more interested into buying an EV. Some technology enthusiast found in a survey that 17% respondents marked "Less CS" to be the greatest concern. However, large numbers of CS would come into existence with large number of EVs on the road. Again, fast charging technology is not feasible with this low rate of EV in the road.

It was also learned that government incentives like free parking and cheaper EV spare parts would increase the possibility of EV-adoption. In addition to this, a study shows that 1 CS for every 100,000 residents could have significant impact on EV-adoption rate. Based on some surveys done about EV general knowledge, it was learned that most of the people hold wrong idea regarding the EV purchases. The respondents impose their economic affairs whereas they are misinformed about a price of an EV. Overall, 2/3rd of the respondents marked wrong information about an EV. Almost, 95% of the people were not aware of the incentives in their locality provided by the government for using an EV. Therefore, raising awareness and correcting the misconception would increase the willingness of consumers to prefer an EV.

The impact of policy incentives towards EVs were studied in detail in [9]. The authors discussed different policy incentives that may attract the adoption of EVs. It is mentioned that a particular policy incentive is effective if the number of EVs purchased rises after the incentive is introduced. The effectiveness is measured by the difference between the number of EVs sold when the policy incentive is applied and that without the policy incentive. They found that incentives to parking (free/paid), public charging (free/paid), access to bus lanes, price including subsidies and range play a major role in the purchase of EVs. The results obtained from the analysis conducted in [9] show that there is a great impact on the willingness of the people to purchase EVs when the above-mentioned incentives are provided to the EV users.

As seen in the previous studies, most of them involved general consumer perspectives and the authors realized that it will be more important to have two different perspective—(i) the perspective of people who are aware of the EV technology and can be considered as technology reviewers, researchers (technical respondents) and (ii) the perspective of people who will be sole user of EV and may or may not be aware of the EV technology (non-technical respondents). This approach may help in providing unique perspective and concluding remarks which were not there in the previous studies. Magnossun in [40], have found that different categorical perspective in surveys, have helped in providing meaningful insights and suggestions and this study also helped in motivating the authors to take similar approach. Present economy of transportation in Qatar is heavily based on fossil fuel. According to the Qatar Vision of 2030, $CO_2$ emission must be reduced by 17% [41]. To achieve this, the Green vehicle Initiative was launched in 2017. The aim of this initiative is to ensure 10% of total cars in Qatar must be EVs by 2022. To achieve this vision, the culture of EVs or hybrid cars must be promoted in this region. Currently, there are 7 EV CS and recently a solar powered EV CS [42] is installed, which are also shown in Figure 2.

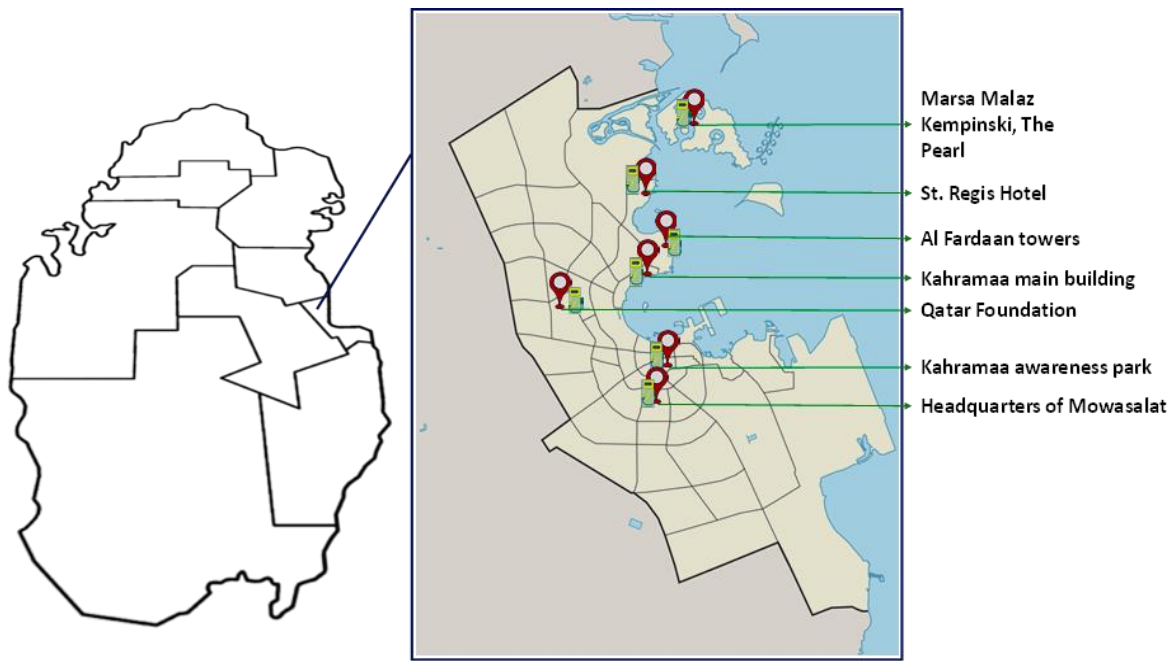

**Figure 2.** EV-charging stations in Doha, Qatar.

## 3. Materials and Methods

With the help of the literature review done on similar studies done in various parts of the work, the purpose of this study is to conduct a survey on different categories of people who have had different understanding and willingness in purchasing EVs in the future. The methodology adopted in the study is shown in Figure 3. The questionnaire begins with questions that ask about the respondents' personal information such as age, gender, nationality and their highest education degree (details in Table 3). An anonymous survey was carried out among local people (technical and non-technical) in Qatar. Instructions were provided mentioning that the study was to analyze the feedback of the respondents about the public perspectives on EVs and their willingness in purchasing those vehicles. The questionnaire was prepared carefully to avoid repetitive question. Authors' previous expertise in preparing surveys were used so that the obtained results can be analyzed for meaningful conclusions [43,44]. Extreme care was taken to ensure that the anonymity of the study and confidentiality of responses were maintained by not asking about the identification of the participants. Furthermore, the data were analyzed and reported in a cumulative manner to prevent the identification of participants. Based on the prominent researches in the field of EV [45,46], the questions were designed so that analysis using two-sample *t*-test can provide meaningful conclusions. The purpose of this study was to analyze if different characteristics of the respondents would have different views on the adoption of EVs. We hypothesized that people who have higher educational degrees would have a better image on EVs and are more likely to purchase them due to their higher awareness on global warming. After that, a question about the number of years of driving experience was asked. Before the questions that are specific about EVs were asked, their willingness to contribute to reduce global warming was asked. It was required to see if there are still people who are unaware or do not care about the effects of global warming (Table 3).

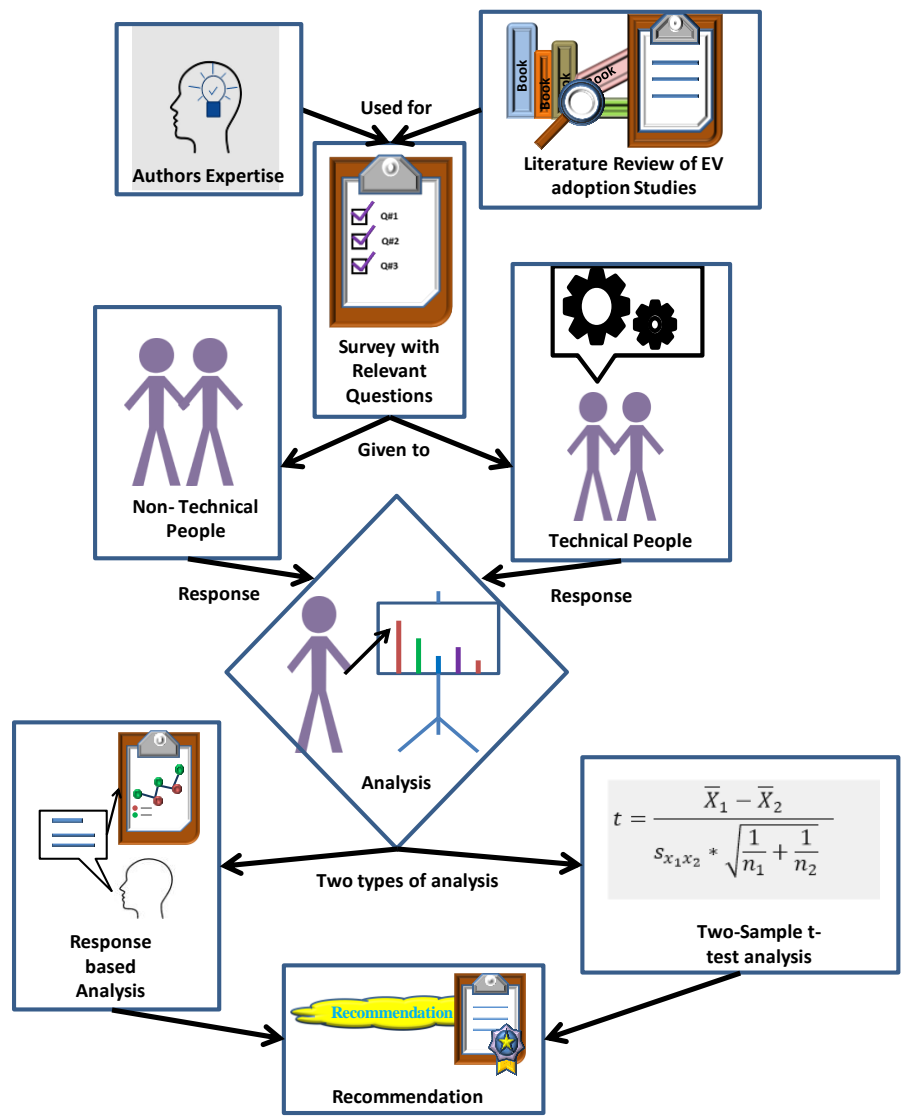

**Figure 3.** Block diagram of the methodology adopted in the study.

Table 3. Survey questions.

| Question Objective | Question Statement |
|---|---|
| General information of the respondents | • Please specify your highest degree of education:<br>• How many years of driving experience have you had?<br>• How many hours do you use your car and how many kilometers do you drive per day? |
| Interest on climate change | • Studies show that climate change leads to serious catastrophes. How willing are you to bring down the climate change?<br>• Studies show that electric vehicles (EV) are more environment-friendly, but it has higher purchasing price than normal cars. How willing are you to purchase an EV? |
| Financial barrier | • The maintenance and charging costs of an EV is 50% cheaper, but a fully charged EV has less driving range than normal cars. How willing are you to prefer an EV? |
| technological barrier | • Considering all the advantages, how willing are you to prefer an EV if it takes 1 h to get fully charged?<br>• If you purchase an EV, what is the reasonable duration to adequately charge an EV? |
| Policy barrier | • How willing are you to purchase an EV if the government provides some incentives such as free parking and discounted spare parts for EVs only? |
| Infrastructure barrier | • If you are an EV owner, select the most convenient location of an EV-charging station.<br>• Do you understand that plugging an EV every time you are parked can lead to lower costs of electric energy, since unlike liquid fuel low power electric energy can be much cheaper than high-power quick charge electric energy?<br>• Would you buy an EV if there were low power automatic charging available, so that your vehicle is on a low charge every time it is parked? |
| Summary response after going through the whole survey | • If EV has higher purchasing price than normal cars, shorter driving range, long charging time, fewer charging stations and no incentive from the government, how willing are you to prefer an EV? |

All the details of the questions in the questionnaire are shown in the Appendix A. Moreover we have divided the questions to get the respondents viewpoints on various known factors affecting the EV-adoption such as Financial, technological, Policy and Infrastructure. The final question in the survey was to summarize if the respondents after going through all the advantages and disadvantages are still willing to purchase an EV. The details of the questions and the objectives (financial barrier, technological barrier, policy barrier, infrastructure barrier, etc.) they correspond to are in Table 3 below.

The purpose of question under interest on climate change was to see if the initial cost of EVs would be one of the major barriers of their adoption despite it being a green mode of transportation. The people who were willing to contribute to global warming reduction may not be willing to purchase EVs after knowing that they have high purchasing costs. Questions on financial barrier were aimed to back up the high initial cost of EVs by informing the people that EVs can be a good investment. However, their shorter driving ranges were mentioned to see if the driving ranges would also be a crucial barrier for EVs. The importance of the EV-charging time could be analyzed by asking questions on technological barrier. It was expected that it would be very unlikely for people to be willing to purchase EVs if the time of charging were still relatively long. The reasonable charging times in order for people to be willing to purchase EVs were also asked in this section and it would be very likely that everyone would go for the fastest possible time.

After showing the advantages and disadvantages of EVs and asking about peoples' willingness due to those factors, questions about EV infrastructures were asked. The respondents were asked how placing the charging stations in different places (i.e., public, work, homes and highways) would affect their decision in purchasing an EV. The importance of providing good incentives for EV users were also asked to convince further the respondents to be more willing in adopting EVs. Then, some points that could prevent people from buying EVs were shown and they were asked if they would purchase EVs due to those factors. This is to see which factors would be the most to the least crucial factors for EV-adoptions. More details of the respondent list such as education background and driving experience (questions in general information of the respondents) can be seen summarized in Figure 4.

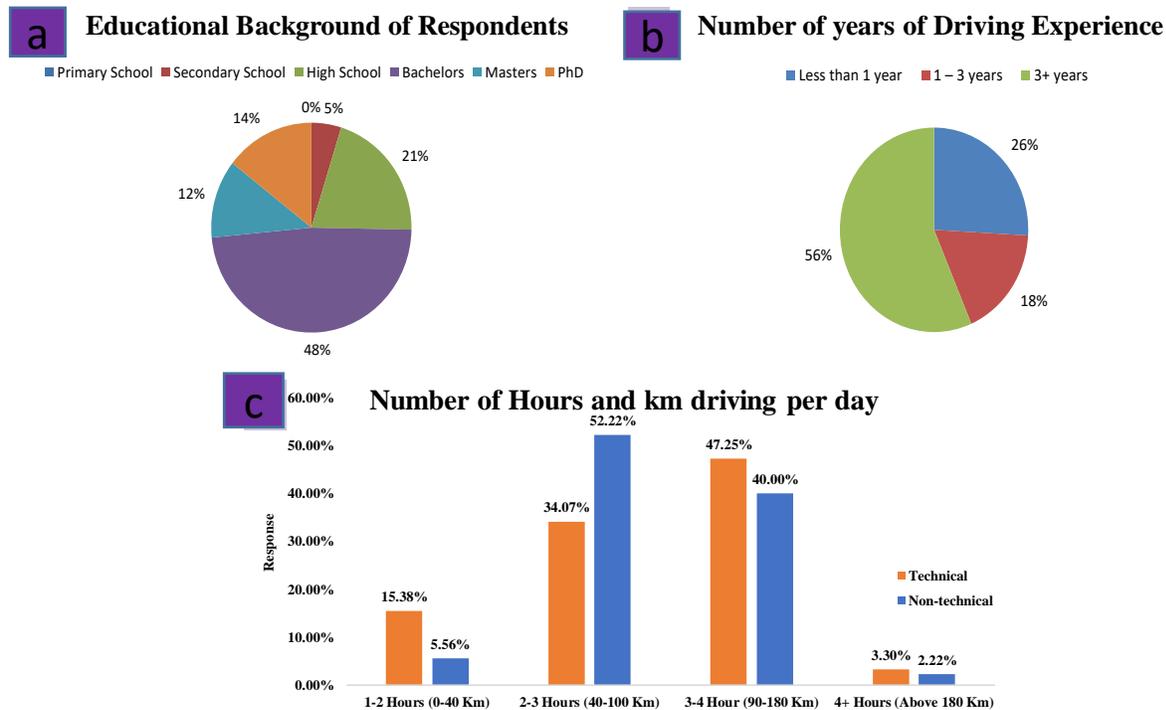

**Figure 4.** Summary of (**a**) Educational Background of the respondents, (**b**) Years of driving experience of the respondents and (**c**) Number of hours and km of driving per day of the respondents.

The respondents were divided into two categories as discussed earlier: technical and non-technical. The technical respondents (95) were from organizations, which were directly or indirectly related to technical fields of EV. The respondents were working in one of the prominent car distributions organization in Qatar and higher educational organization in the state of Qatar. The non-technical respondents (88) were local people working in non-technical fields such as banks, medical organizations, schools and non-profit organizations.

## 4. Analyses

The analyses of data was done in this study in two major ways: (i) analyzing the response-based indicator, i.e., the percentage in each answer category for each question, (ii) Two-sample *t*-test statistics to analyze whether the responses from both the groups are significantly different or not.

Response-Based Indicator

The percentages of answer in each category for different questions were analyzed to get a perception of most of the respondents. These results were analyzed and recommendations were obtained based on this.

T-Statistics

In a statistical analysis, *t*-statistics is used to compare the means of two different groups. It is also called the *t*-test theory while applying it on survey results. Anyone can identify two different responses, but to find if the difference is statistically significant or not, can be done with the help of *t*-test theory. *t*-test theory works very well with survey database. As it compares and determines statistical difference, survey results from two groups are very good data to analyze with *t*-test theory. Based on the types of the samples and groups, *t*-test can be done in three different ways: (i) one-sample *t*-test, (ii) two-sample *t*-test, and (iii) paired *t*-test. For the purpose of analysis in this study, we will be using the two-sample *t*-test analysis which works with the means of two independent groups' response and finds if the means are significantly different from each other or not, by comparing the *t*-statistics value with the critical value. A *t*-test can only distinguish if the means are significantly different; it is the user's decision to make a meaning conclusion from this difference.

Each question in the survey which were analyzed using the *t*-statistics had 5 options which were coded from 1 to 5, for example—strongly unwilling was coded as 1, unwilling was coded as 2, neutral was coded as 3, willing was coded as 4 and strongly willing was coded as 5, can be seen in Figure 5.

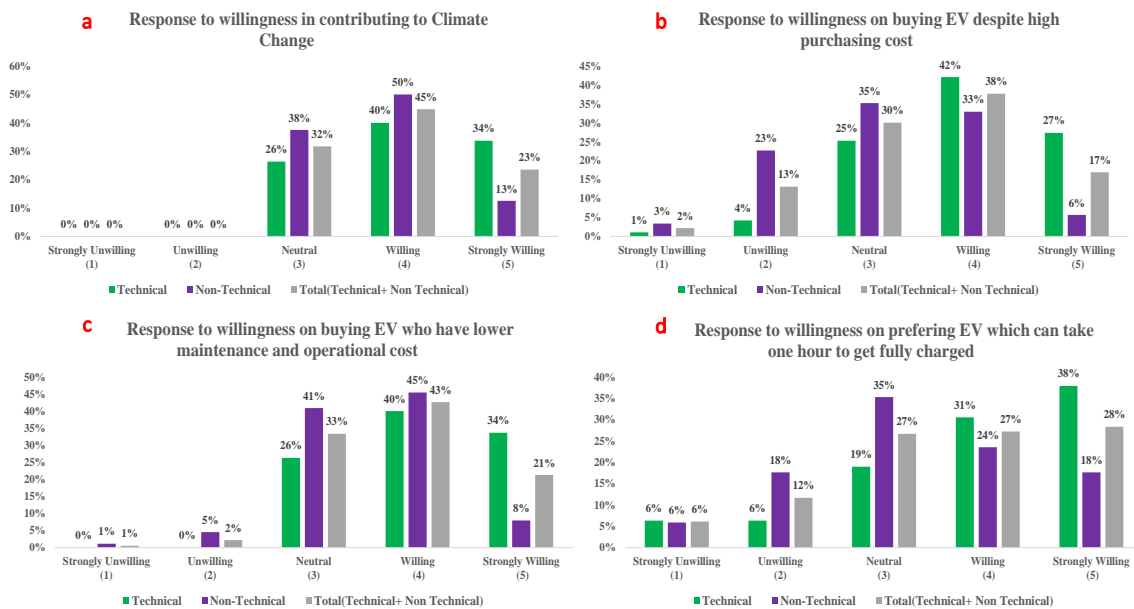

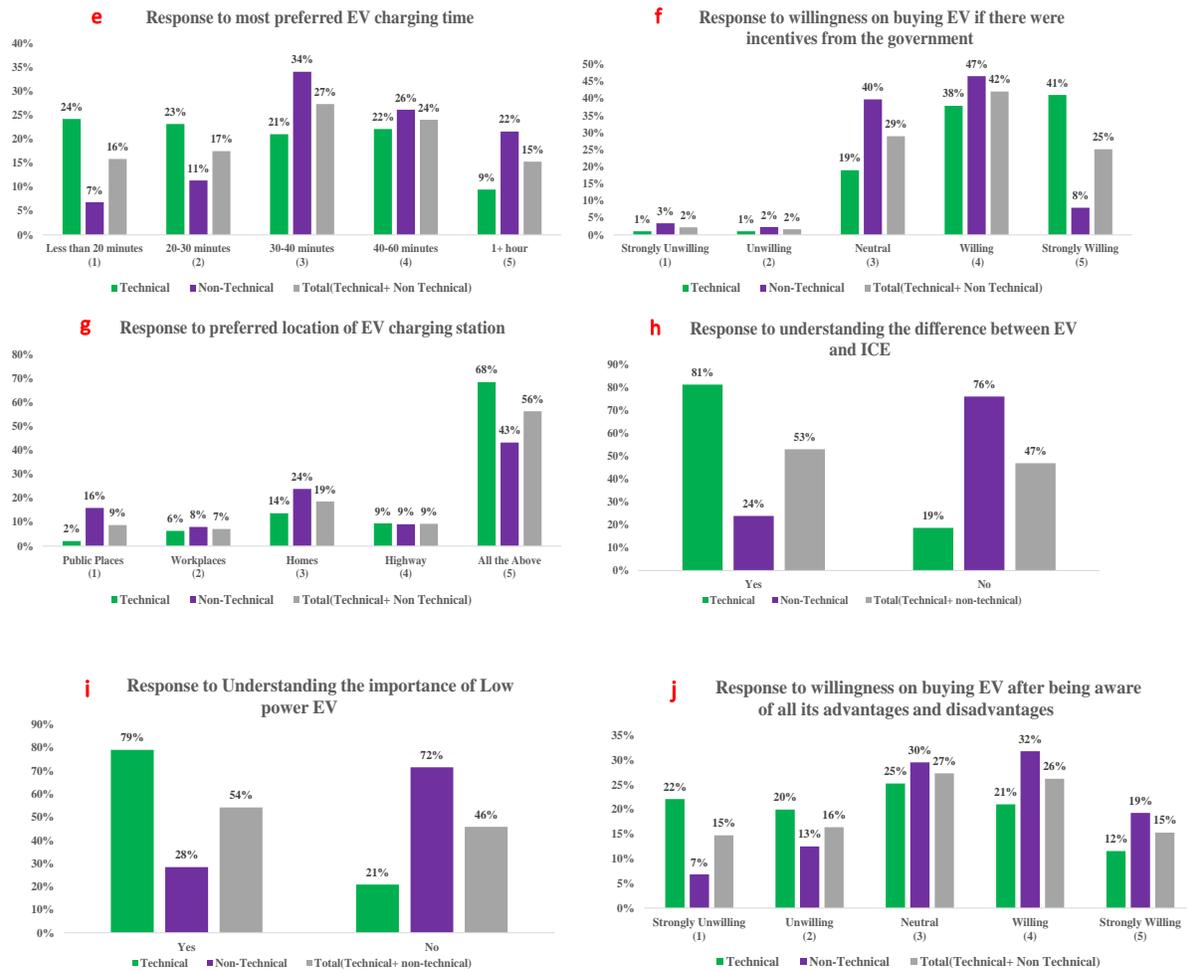

**Figure 5.** Bar chart representation of the responses to questions in Table 3. (**a**) Response to willingness in contributing to Climate Change, (**b**) Response to willingness on buying EV despite high purchasing cost, (**c**) Response to willingness on buying EV who have lower maintenance and operational cost, (**d**) Response to willingness on prefering EV which can take one hour to get fully charged, (**e**) Response to most preferred EV charging time, (**f**) Response to willingness on buying EV if there were incentives from the government, (**g**) Response to preferred location of EV charging station, (**h**) Response to understanding the difference between EV and ICE, (**i**) Response to understanding the importance of low power EV, (**j**) Response to willingness on buying EV after being aware of its advantages and disadvantages.

5. Results and Discussion

In this section, the results from the survey are summarized and comparisons between technical, non-technical respondents and the two categories combined are shown in the bar graphs Figure 5a–j for the questions in the interest on climate change objective onwards in Table 3.

As stated in the earlier section two-sample *t*-test statistics were done on the responses between the two categories of respondents (technical and non-technical) to see if they are significantly different from each other. The combined results are also shown in figure to see the approach in the study provides conclusive results. The results are shown in Table 4 and the details of how it was conducted were stated in the analysis section.

**Table 4.** Two-sample *t*-test results on the answers to the survey questions.

| Questions | Two-Sample *t*-test Analysis Result | Conclusion That Can Be Drawn |
|---|---|---|
| Studies show that climate change leads to serious catastrophes. How willing are you to bring down the climate change? | Mean of technical respondents = 4.07<br>Mean of non-technical respondents = 3.75<br>By two-sample *t*-test analysis the responses are significantly different from each other. | The technical people understand the impact of climate change and are more willing to contribute unlike the non-technical people who are not as strongly willing. This can also be understood from the fact that technical people can contribute more to bringing down the climate change with their knowledge and expertise. |
| Studies show that electric vehicles (EVs) are more environment-friendly, but it has higher purchasing price than normal cars. How willing are you to purchase an EV? | Mean of technical respondents = 3.91<br>Mean of non-technical respondents = 3.15<br>By two-sample *t*-test analysis the responses are significantly different from each other. | The technical people are more willing to buy EVs. The non-technical also show positive attitude towards buying EVs, but the responses are not similar to technical people. This can also be confirmed from the fact that the non-technical people are still not fully aware of Green Mode of transportation (Previous Question on the interest on climate change). |
| The maintenance and charging costs of an EV is 50% cheaper, but a fully charged EV has less driving range than normal cars. How willing are you to prefer an EV? | Mean of technical respondents = 4.07<br>Mean of non-technical respondents = 3.55<br>By two-sample *t*-test analysis the responses are significantly different from each other. | The technical people are more willing to buy it looking at cost saving due to maintenance compared to less driving range. The non-technical people are also willing, but do not have similar response like the technical people. This means there needs to be more communication among the general public to inform the long term savings in terms of operational and maintenance cost when using EVs. |
| Considering all the advantages, how willing are you to prefer an EV if it takes 1 h to get fully charged? | Mean of technical respondents = 3.87<br>Mean of non-technical respondents = 3.29<br>By two-sample *t*-test analysis the responses are not significantly different from each other. | Both the groups have similar response to buying EVs with the requirement of 1 h for getting fully charged. This is interesting where the responses of technically knowledgeable group have similar response with the Non-technical people. |
| If you purchase an EV, what is the reasonable duration to adequately charge an EV? | Mean of technical respondents = 2.69<br>Mean of non-technical respondents = 3.44<br>By two-sample *t*-test analysis the responses are not significantly different from each other. | With the previous question being similar, it is no surprise for this question the responses are not significantly different, which is interesting as the non-technical people have similar expectation for the charging time than the fully aware technical people. Thus, awareness among the non-technical people can help in easier adoption of the EV as the expectations are not illogical. |
| How willing are you to purchase an EV if the government provides some incentives such as free parking and discounted spare parts for EVs only? | Mean of technical respondents = 4.17<br>Mean of non-technical respondents = 3.53<br>By two-sample *t*-test analysis the responses are significantly different from each other. | The technical people are more aware of how the government incentives can help in adopting EVs whereas the non-technical people are confused of how the incentives can be useful and thus responded mostly as neutral. |
| If you are an EV owner, select the most convenient location of an EV-charging station. | Mean of technical respondents = 4.36<br>Mean of non-technical respondents = 3.56<br>By two-sample *t*-test analysis the responses are not significantly different from each other. | Both the groups believe in having the charging stations in all locations. |
| If EVs have higher purchasing price than normal cars, shorter driving range, long charging time, fewer charging stations and no incentive from the government, how willing are you to prefer an EV? | Mean of technical respondents = 2.8<br>Mean of non-technical respondents = 3.44<br>By two-sample *t*-test analysis the responses are not significantly different from each other. | The willingness of both the groups to purchase an EV after being told of all the advantage and disadvantage of EVs has increased. |

As expected the combined results could have the important information regarding the lack of awareness regarding EVs in general public and can be verified from Figure 5h, (i) where were questions on understanding the difference between EVs and ICE vehicles and understanding the importance of low-power EVs, respectively. Similar comparison could be done in Figure 5a–c where the strong willingness of the technical people to contributing to climate change, buying EVs despite high purchasing cost is evident compared to the non-technical people, and this could have missed if the different perception were not captured. In figures, looking at combined results one could have easily missed the important conclusion that there is lack of awareness about sustainable mode of transportation among the non-technical people. It is evident from the results in Figure 5 and Table 4 that the general public needs to be made more aware regarding the green mode of transportation such as electric vehicles. A majority of the non-technical people was in favor of buying EVs, despite higher initial cost. There were some technical people who were neutral to this question and thus can be assumed that they believe that the prices of EVs can be made lower through research. In the following question where the respondents were informed regarding the low maintenance and operational cost of EVs, the technical people were in majority very positive in buying EVs unlike the non-technical people, but it can be confirmed from Figure 5c that the financial barrier is not a major hindrance to the easy adoption of EVs in Qatar. The response to the question on longer charging time of EVs were not significantly different between the two groups and can be used to assume that awareness for greener mode of transportation despite its lacking can be promoted as they do not have too illogical expectations. This can be considered as a positive sign as the unaware non-technical group were answering similar to the aware technical group and thus people are ready for the various options of charging time. The question on how government incentives can help in EV-adoption had technical people responding positive unlike the non-technical, which can be attributed to their less awareness on how such incentives can help in quick adoption of such green and sustainable mode of transportation. All the respondents replied similarly in preferring charging stations on all locations. There was a clear difference in the response to the question on understanding the difference between EVs and ICE vehicles and the preference of low power EV, which will help in charging in all places. As expected the technical people responded 'Yes' to these questions in majority unlike the non-technical group. Finally, a good fraction of both the categories were willing to buy EVs after understanding the advantages and disadvantages of EVs.

## 6. Conclusions

The adoption of EVs is no doubt one of the best solutions for a greener environment, due to their low emissions of GHGs relative to ICE vehicles. However, barriers such as long charging times, lack of policy incentives, initial cost and insufficient charging infrastructures are significantly slowing down the widespread adoption of EVs globally. This study has investigated the different perspectives of technical and non-technical people on the adoption of EVs in Doha, Qatar, where almost all the existing vehicles are fossil fuel dependent. Survey questions were distributed and analyzed using two methods; (i) response based analysis and (ii) two-sample *t*-test analysis. The analysis has shown that unlike many other countries financial barrier is a not a major hindrance to the adoption of EVs but there needs to more awareness among the general public about climate change and how it can be tackled with greener and sustainable mode of transportation such as EVs. The people when made aware will help in quicker adoption of EVs which can be further boosted by adequate government incentives. Government incentives can play a major role in EV-adoption, as seen from the survey results and also proved from the successful adoption in countries like Norway. From the results obtained, the top factor that would prevent people from purchasing EVs is the lack of public awareness policy incentives towards EVs. Other potential EV adoption barriers were also informed to the people such as the longer charging time and the higher purchase cost but these do not have a significant effect on the peoples' willingness in purchasing EVs as long as they are within a suitable range. The shorter driving range of EVs was also not a major problem, since most people were still willing to purchase EVs, while knowing that the maintenance costs of EVs were cheaper than ICE vehicles. This may be due to the fact that Qatar is a small country with many fueling stations that could be integrated with EV-charging stations, so short driving ranges would not be a barrier to adoption. The authors believe that developing a mobile application that could share the latest features and news of EVs could help to raise awareness among the general public. This mobile application could be used in the future to suggest EV-charging stations to users as well to reserve slots—thus making EV-charging even more convenient for them and helping in the gradual adoption of EVs. Without a doubt, the number of EVs in Qatar will increase exponentially if the incentives towards EVs such as free parking in all places, cheaper spare parts and the selling of electricity using the vehicle-to-grid (V2G) system becomes available to consumers. While providing EV users with many incentives, providing them with sufficient charging infrastructures in various places is also an important solution for EV-adoption. Most of the respondents want charging stations to be available in multiple places (i.e., public places, workplaces, homes and highways), so providing EV users with convenient accessibility to charging infrastructures will convince more people to switch to EVs.

**Author Contributions:** A.K., A.I., M.E.H.C. and S.M.A.U.Z were involved in the conceptualization of the study. A.R., A.A.A.B. and M.R.A. had conducted the survey and the analysis of the study. All the authors were involved in the drafting of the study and the analysis involved in the study.

**Funding:** This research received no external funding.

**Acknowledgement:** This publication was made possible by the UREP grant # [24-091-2-018] from the Qatar National Research Fund (a member of the Qatar Foundation). The statements made herein are solely the responsibility of the authors. The authors would also like to thank the Qatar National Library (QNL) who has helped in the publication of the article.

**Conflicts of Interest:** The authors declare no conflict of interest.

**Appendix A**

## Perspective of EVs

**1. Please specify your highest degree of education:**

○ Primary School  ○ Bachelors

○ Secondary School  ○ Masters

○ High School  ○ PhD

**2. How many years of driving experience have you had?**

○ Less than 1 year

○ 1-3 years

○ 3+ years

**3. How many hours do you use your car and how many kilometres do you drive per day?**

○ 1-2 Hours (0-40 Km)

○ 2-3 Hours (40-100 Km)

○ 3-4 Hours (90-180 Km)

○ 4+ Hours (Above 180 Km)

4. Studies show that climate change leads to serious catastrophes. How willing are you to bring down the climate change?

○ Strongly Unwilling          ○ Willing

○ Unwilling                   ○ Strongly Willing

○ Neutral

5. Studies show that Electric Vehicles (EVs) are more environment-friendly but it has higher purchasing price than normal cars. How willing are you to purchase an EV?

○ Strongly Unwilling          ○ Willing

○ Unwilling                   ○ Strongly Willing

○ Neutral

6. The maintenance and charging costs of an EV is 50% cheaper but a fully charged EV has less driving range than normal cars. How willing are you to prefer an EV?

○ Strongly Unwilling          ○ Willing

○ Unwilling                   ○ Strongly Willing

○ Neutral

7. **Considering all the advantages, how willing are you to prefer an EV if it takes 1 hour to get fully charged?**

- ◯ Strongly Unwilling
- ◯ Unwilling
- ◯ Neutral
- ◯ Willing
- ◯ Strongly Willing

8. **If you purchase an EV, what is the reasonable duration to adequately charge an EV?**

- ◯ Less than 20 minutes
- ◯ 20-30 minutes
- ◯ 30-40 minutes
- ◯ 40 minutes - 1 hour
- ◯ 1+ hour

9. **How willing are you to purchase an EV if the government provides some incentives such as free parking and discounted spare parts for EVs only?**

- ◯ Strongly Unwilling
- ◯ Unwilling
- ◯ Neutral
- ◯ Willing
- ◯ Strongly Willing

10. **If you are an EV owner, select the most convenient location of an EV charging station**

- ◯ Public Places (e.g. Malls, Parks, etc)
- ◯ Workplaces (e.g. Offices, Industries, etc.)
- ◯ Homes
- ◯ Highway Roads
- ◯ All the Above

**11. Do you understand that plugging an EV every time you are parked can lead to lower costs of electric energy, since unlike liquid fuel low power electric energy can be much cheaper than high-power quick charge electric energy?**

○ Yes

○ No

**12. Would you buy an EV if there were low power automatic charging available, so that your vehicle is on a low charge every time it is parked?**

○ Yes

○ No

**13. If EV has higher purchasing price than normal cars, shorter driving range, long charging time, fewer charging stations and no incentive from the government, how willing are you to prefer an EV?**

○ Strongly Unwilling

○ Unwilling

○ Neutral

○ Willing

○ Strongly Willing

**Figure A1.** Final Survey provided to the respondents